\def\la{\mathrel{\hbox{\rlap{\hbox{\lower4pt\hbox{$\sim$}}}\hbox{$<$}}}}
\def\ga{\mathrel{\hbox{\rlap{\hbox{\lower4pt\hbox{$\sim$}}}\hbox{$>$}}}}

\def\Teff{\ifmmode{T_{\rm eff}}\else{\hbox{$T_{\rm eff}$} }\fi}
\def\Rzero{\ifmmode{R_0}\else{\hbox{$R_0$} }\fi}
\def\kms{km s$^{-1}$}

\def\56ni{$^{56}$Ni}
\def\56co{S^{56}$Co}

\def\dl{\Delta\lambda}

\newcount\counttemp
\newcount\counteqn
\def\eqnnumber#1){\number\counteqn \rm #1)}
\def\eqn#1){\global\advance\counteqn by 1   \eqnnumber \rm #1)}
\def\aeqn#1){\global\advance\counteqn by 1   \eqnnumber \rm #1)}
\def\showeqn#1){\counttemp=\counteqn \advance\counttemp by #1
                \number\counttemp)}
\def\showeqnsq#1]{\counttemp=\counteqn \advance\counttemp by #1
                \number\counttemp]}
\def\showeqnx#1,#2){\counttemp=\counteqn \advance\counttemp by #1
                \number\counttemp#2)}

\documentclass{aastex}

\received{}
\accepted{}
\journalid{}{}
\articleid{}{}
\shorttitle{Supernova Resonance--Scattering Profiles}
\shortauthors{Branch \& Jeffery}

\begin{document}
\title {Supernova Resonance-Scattering Profiles in the Presence of
External Illumination}

\author {{David Branch\altaffilmark{1}}, {David J.~Jeffery\altaffilmark{2}}, {Myra
Blaylock\altaffilmark{1}}, and {Kazuhito Hatano\altaffilmark{1}}}

\altaffiltext{1}{Department of Physics and Astronomy, University of
Oklahoma, Norman, Oklahoma 73019, USA: branch@mail.nhn.ou.edu}

\altaffiltext{2}{Department of Physics and Astronomy, Middle Tennessee
State University, 1301 East Main St., Murfreesboro, TN 37132}

\begin{abstract}

We discuss a simple model for the formation of a supernova spectral
line by resonance scattering in the presence of external illumination
of the line-forming region by light from circumstellar interaction
(toplighting).  The simple model provides a clear understanding of the
most conspicuous toplighting effect: a rescaling or, as we prefer, a
``muting'' of the line profile relative to the continuum.  This effect
would be present in more realistic models, but would be harder to
isolate.  An analytic expression for a muting factor for a P-Cygni
line is derived that depends on the ratio $E$ of the toplighting
specific intensity to the specific intensity from the supernova
photosphere.  If $E<1$, the line profile is reduced in scale or
``muted''.  If $E=1$, the line profile vanishes altogether.  If $E>1$,
the line profile flips vertically: then having an absorption component
near the observer-frame line center wavelength and a blueshifted
emission component.

\end{abstract}


\section{Introduction}

In a simple but useful model of spectral line formation during the
photospheric phases of a supernova, a line forms by resonance
scattering in a homologously expanding atmosphere above a sharp
photosphere.
An unblended line has a P-Cygni profile with an emission feature
near the observer-frame line center wavelength and a blueshifted absorption.
Synthetic spectra calculated on the basis of this simple model
have been found to fit the observed spectra of most supernovae
reasonably well, and to be useful for establishing constraints on the
composition structure of the ejected matter.  For a detailed
discussion of the model, see Jeffery \& Branch (1990), who also
provided an atlas of line profiles, and compared synthetic spectra to
numerous observed spectra of the Type~II SN~1987A.
Recent studies based on the simple model, using the SYNOW supernova
synthetic-spectrum code,
include Millard et~al. (1999) on the Type~Ic SN~1994I and Hatano
et~al. (1999) on the Type~Ia SN~1994D.

In the bright Type~IIn SN~1998S we have encountered an event for which
the above simple model fails.  The supernova SN~1998S has been
observed extensively with the {\sl Hubble Space Telescope (HST)} as a
target of opportunity by the Supernova INtensive Study (SINS) group
(Garnavich et~al. 1999, Lentz et~al. 1999, and Blaylock et~al. 1999,
all in preparation), as well as from the ground (Leonard et~al. 1999).
The observed spectra, especially at short wavelengths and early
epochs, contain numerous narrow ($\la500$~\kms) absorption and
emission features that formed in circumstellar matter.  The optical
and ultraviolet spectra also contain broad ($\sim5000$--10,000~\kms)
features that formed in the supernova ejecta.  All of the broad
features show little contrast with the continuum, especially at early
times.  It is not plausible that the radial dependence of the optical
depths of all of the supernova lines should be such that the line
profiles come out to be shallow, and in synthetic spectrum
calculations with the SYNOW code in its simplest form we cannot
account for the relative strengths of the spectral features of
SN~1998S.  The presence of the circumstellar features suggests that
the broad supernova features were being affected by light from the
region of circumstellar interaction: i.e., the line formation region
was illuminated not only from below by light from the photosphere, but
also from above by light from the circumstellar interaction.  We refer
to this external illumination of the supernova line-forming region as
``toplighting''.

In this paper we present a simple model of
resonance-scattering line formation in the presence of toplighting.
This simple model provides a clear understanding
of the most conspicuous toplighting effect:  a rescaling or, as we prefer
to call it, a ``muting'' of the line
profiles relative to the continuum.
This effect would be present in more realistic models, but would be
harder to isolate.

Section~2 presents the model.
The emergent specific intensities are derived in \S~3.
The line profile and a muting factor are derived in \S~4.
In \S~5, we offer a picturesque description of radiative transfer
in the model atmosphere.
A final discussion appears in \S~6.

\section{The Model}

Suppose that supernova line formation takes place by isotropic resonance
scattering in a spherically symmetric atmosphere that has a sharp
photosphere with radius $R_{\rm ph}$ as an inner boundary and
a circumstellar-interaction region (CSIR) as an outer boundary.
Assume that the CSIR is a shell of zero physical width and optical depth,
and let its radius be $R_{\rm cs}$ (see Fig.~1):  $R_{\rm cs}\geq R_{\rm ph}$
in all cases, of course.
For simplicity, assume
that the photosphere emits an outward angle-independent specific intensity
$I_{\rm ph}$, that the CSIR emits isotropically with specific
intensity $I_{\rm cs}$,
and that $I_{\rm ph}$ and $I_{\rm cs}$ are constant over the wavelength
interval of interest.
When $I_{\rm cs}$ is set to zero, we have what we will call the standard
case.
Nonzero $I_{\rm cs}$ gives the toplighting case.

     We assume that the atmosphere is in homologous expansion.
In homologous expansion the radius of a matter element $r$ is given by
$$  r=vt        \,\,  ,   \eqno(\eqn)$$
where $v$ is the constant radial velocity of the matter element and $t$ is
the time since explosion which is assumed sufficiently large that initial
radii are negligible.

Line formation is treated by the (non-relativistic) Sobolev method
(e.g., Rybicki \&~Hummer 1978;
Jeffery \&~Branch 1990;  Jeffery \&~Vaughan 1999) in which the line profile
in the emergent flux spectrum depends on the radial behavior of the Sobolev
line optical depth $\tau$ and the line source function $S$ (which is
a pure resonance scattering source function following our earlier assumption). 
We consider only isolated (unblended) line formation
and do not include any continuous opacity in the atmosphere.

      We note for homologous expansion that the resonance surfaces
for observer-directed beams are planes perpendicular to the line-of-sight.
(A resonance surface is the locus of points on which beams are
Doppler shifted into resonance with a line in an atmosphere with
a continuously varying velocity field.)
If we take $z$ to be the line-of-sight coordinate with the positive
direction toward the observer, the resonance plane for a wavelength
shift $\Delta\lambda$ from the line center wavelength $\lambda_{0}$ is at
$$  z=-{ {\Delta\lambda} \over { \lambda_{0} } }ct   \,\,  ,    \eqno(\eqn)$$
where $z/t$ is the plane's velocity in the $z$-direction.
Blueshifts give positive $z$ planes and redshifts, negative $z$ planes.

\medskip

\section{The Emergent Specific Intensities}

For resonance scattering the source function of a
line is just equal to the mean intensity.
In our model the line source function is
$$ S(r) = WI_{\rm ph} + [1-W]I_{\rm cs} \,\, , \eqno(\eqn)$$
where W is the usual geometrical dilution factor,
$$ W = {{1}\over{2}}
\left[1-\sqrt{1-\left({R_{\rm ph}}\over{r}\right)^{2}}\,\right]
                                                \eqno(\eqn) $$
(e.g., Mihalas 1978, p.~120).
The first term in equation~(\showeqn -1) accounts for radiation from the
photosphere and the second, for radiation from the CSIR. 

Consider a line-of-sight specific intensity beam that does not intersect the
photosphere (i.e., one that has impact parameter $p > R_{\rm ph}$ in
the standard $p, z$ coordinate system) and that has a wavelength in
the observer's frame that differs from the line center wavelength by
$\Delta\lambda$.  In a toplighting case such a beam originates from
the CSIR as shown in Figure~1.  From the Sobolev method, the emergent
specific intensity of such a beam can be seen to be
$$ I_{\dl} (p>R_{\rm ph}) = I_{\rm cs}\ e^{-\tau} +[W I_{\rm ph} +
(1-W) I_{\rm cs}]\ (1 - e^{-\tau}) + I_{\rm cs} \,\, , \eqno(\eqn) $$
where, of course, $W$ and $\tau$ must be evaluated at the
Sobolev resonance point for the $\dl$ value under consideration.
Similarly, the emergent specific intensity of a line-of-sight
beam that originates from the photosphere
(i.e., has $p\leq R_{\rm ph}$) can be seen to be
$$ I_{\dl} (p\leq R_{\rm ph}) = I_{\rm ph}\ e^{-\tau} + [W I_{\rm ph}
                        + (1-W) I_{\rm cs}]\ (1 - e^{-\tau}) + I_{\rm cs}
                                                   \,\, .    \eqno(\eqn) $$

Equations~(\showeqn -1) and~(\showeqn 0) can be rearranged into
the convenient expressions
$$ I_{\dl} (p>R_{\rm ph}) = (I_{\rm ph} - I_{\rm cs})W(1 - e^{-\tau})
                 + 2 I_{\rm cs}              \eqno(\eqn) $$
and
$$ I_{\dl} (p\leq R_{\rm ph}) = (I_{\rm ph} - I_{\rm cs})e^{-\tau}
                   + (I_{\rm ph} - I_{\rm cs})W(1 - e^{-\tau})
                   + 2 I_{\rm cs} \,\, .         \eqno(\eqn) $$
If $I_{\rm cs}=0$, these expressions reduce to the standard or non-toplighting
case Sobolev expressions for emergent specific intensity.
If the resonance point is not in the line-forming region (i.e., it occurs
at $r<R_{\rm ph}$ or $r > R_{\rm cs}$,
or it is in the occulted region from which no
beam can reach the observer), then $\tau$ is just set to zero in the
expressions which then reduce to the continuum expressions
$$ I_{\dl} (p>R_{\rm ph}) =  2 I_{\rm cs}                        \eqno(\eqn) $$
and
$$ I_{\dl} (p\leq R_{\rm ph}) = I_{\rm ph} + I_{\rm cs} \,\, .   \eqno(\eqn) $$

     The emission component of a P-Cygni line is largely due to
$I_{\dl} (p>R_{\rm ph})$ beams and the absorption component, to
$I_{\dl} (p\leq R_{\rm ph})$ beams.  To see how going from the
standard to the toplighting case affects the components it is best to
consider equations~(\showeqn -3) and~(\showeqn -4).  From
equation~(\showeqn -3), we see that toplighting tends to reduce the
line emission by changing $I_{\rm ph}$ to $I_{\rm ph} - I_{\rm cs}$.
>From equation~(\showeqn -4), we see that toplighting tends to fill in
the absorption trough (caused by the $e^{-\tau}$ factor of the $I_{\rm
ph}e^{-\tau}$ term) by adding a positive term to the source function.
Both emission component and absorption trough are reduced relative to
the continuum by the addition of continuum terms: $2I_{\rm cs}$ in the
first case and $I_{\rm cs}$ in the second.  Thus we can see (at least
for $I_{\rm ph}>I_{\rm cs}$) that adding toplighting to an atmosphere
is likely to reduce the relative size of line components (i.e., to
mute them).  When $I_{\rm ph}=I_{\rm cs}$, the line components should
vanish altogether as comparing equations~(\showeqn -3)--(\showeqn -2)
and equations~(\showeqn -1)--(\showeqn -0) shows.  In \S~4, we give a
definite analytical analysis of the effect of toplighting on line
profile formation and confirm the muting effect.

\section{The Line Profile and the Muting Factor}

     The flux profile seen by a distant observer is obtained from
$$ F_{\dl} = 2 \pi \int_{0}^{R_{\rm cs}} dp\,p I_{\dl}(p) \,\, ,\eqno (\eqn)$$
where the integration is over impact parameter.
(Actually, the quantity in equation~(\showeqn -0) divided by
the square of the distance to the observer is the flux measured by the
observer.
But for brevity here and below we just call this quantity flux.)
We can obtain semi-analytic expressions for the flux in the standard and
toplighting cases by breaking
the integration for the flux into components.
These semi-analytic expressions then allow a fully analytic formula
for a muting factor which describes the muting effect of toplighting. 

     By inspection of equations~(\showeqn -4)--(\showeqn -0) with
$I_{\rm cs}=0$, we obtain the standard-case results for flux
in the continuum and in the line.
The continuum flux is
$$ F({\rm con})=I_{\rm ph}F_{0}          \,\, , \eqno(\eqn)$$
where
$$ F_{0}=\pi R_{\rm ph}^{2}             \,\, . \eqno(\eqn)$$
The line flux is
$$ F_{\dl}=I_{\rm ph}\left(F_{1}+F_{2}\right) \,\, , \eqno(\eqn)$$
where
$$ F_{1}  = 2 \pi \int_0^{R_{\rm ph}}dp\,p e^{-\tau}  \eqno(\eqn)$$
and
$$ F_{2}  = 2\pi\int_0^{\sqrt{R_{\rm cs}^{2}-z^{2}}}dp\,p
                 W\left(1-e^{-\tau}\right) \,\, .\eqno(\eqn)$$
Note that the $F_{1}$ and $F_{2}$ factors are dependent on $z$ and therefore
on wavelength.
Also recall that $\tau=0$ for a resonance point outside of the atmosphere
or in the occulted region.
The contrast factor
(i.e., relative difference of line flux from continuum flux)
for the standard case is
$$ { {F_{\dl}-F({\rm con})}\over {F({\rm con})}  }
   = { {F_{1}+F_{2}-F_{0}}\over{F_{0}} }         \,\, .\eqno(\eqn)$$

     Again from equations~(\showeqn -10)--(\showeqn -6), but with
nonzero $I_{\rm cs}$, we obtain by inspection the toplighting-case results
for flux in the continuum and line.
The continuum flux is 
$$ F^{\rm top}({\rm con})
 =I_{\rm ph}F_{0}+I_{\rm cs}\left(2G_{\rm 0}-F_{0}\right)
 =\left(I_{\rm ph}-I_{\rm cs}\right)F_{0}+2I_{\rm cs}G_{\rm 0}
                     \,\, , \eqno(\eqn)$$
where
$$ G_{0}=\pi R_{\rm cs}^{2}    \,\, .      \eqno(\eqn)$$
The line flux is
$$ F^{\rm top}_{\dl} =\left(I_{\rm ph}-I_{\rm
cs}\right)\left(F_{1}+F_{2}\right) +2I_{\rm cs}G_{\rm 0} \,\,
. \eqno(\eqn)$$ The $2I_{\rm cs}G_{0}$ terms in equations~(\showeqn
-2) and~(\showeqn -0) just account for the toplighting specific
intensity beams aimed toward the observer from both the near and far
hemisphere of the CSIR.  These beams of course can interact with the
line and the photosphere and this interaction is accounted for in the
other terms.  Note that $G_{0}\geq F_{0}$ and $G_{0}\geq F_{1}$, where
the equalities hold only in the degenerate case where $R_{\rm
cs}=R_{\rm ph}$.  Also note that $G_{0}\geq F_{2}$, where the equality
holds only in the degenerate case where both coefficients are zero:
i.e., $R_{\rm cs}=R_{\rm ph}=0$.  From these inequalities, it is clear
that $F^{\rm top}({\rm con})$ and $F^{\rm top}_{\dl}$ can never be
less than zero: a physically obvious result, of course.

The contrast factor in the toplighting case is
$$
 { { F^{\rm top}_{\dl}-F^{\rm top}({\rm con}) }\over {F^{\rm top}({\rm con})} }
   ={
     {\left(I_{\rm ph}-I_{\rm cs}\right)\left(F_{1}+F_{2}-F_{0}\right)}
     \over
     {\left(I_{\rm ph}-I_{\rm cs}\right)F_{0}+2I_{\rm cs}G_{\rm 0} }
      }
   ={
     {\left(1-E\right)\left(F_{1}+F_{2}-F_{0}\right) }
     \over
     { (1-E)F_{0}+2EG_{\rm 0} }
     }
                                            \,\, , \eqno(\eqn) $$
where
$$ E\equiv{ {I_{\rm cs}}\over{I_{\rm ph}} }   \,\, . \eqno(\eqn)$$

     We define the muting factor $m$ to be the ratio
of the toplighting-case contrast factor to the standard-case one:
$$ m= { {1-E} \over {1-E+2E\left(R_{\rm cs}/R_{\rm ph}\right)^{2} } }
\,\, . \eqno(\eqn)$$ Since all $F_{1}$ and $F_{2}$ factors have
cancelled out, $m$ is fully analytic and wavelength independent.  The
muting factor $m$ is a monotonically decreasing function of $E$ with a
physical maximum of 1 at $E=0$ and with only one stationary point, a
minimum at $E=\infty$.  The muting factor in fact goes to zero at
$E=1$ and becomes negative for $E>1$.  Thus for $E=1$ the P-Cygni
profile of a line vanishes and for $E>1$ the line flips: there is an
absorption feature at the line center wavelength and a blueshifted
emission feature.  (Note that this flipped P~Cygni profile is not the
same as the ``inverse'' P~Cygni profile that would be produced by a
contracting rather than an expanding line-forming region.)  A
picturesque way of understanding a flipped P-Cygni line is given in
\S~5.

     The minimum value of $m$ at $E=\infty$ is given by
$$ m(E=\infty)={{-1}\over{-1+2\left(R_{\rm cs}/R_{\rm ph}\right)^2}  }
                                                          \,\, . \eqno(\eqn)$$
Note that $m(E=\infty)\geq -1$ with the equality holding only for
$R_{\rm cs}/R_{\rm ph}=1$ which is the lower limit on
$R_{\rm cs}/R_{\rm ph}$.
Thus
$$ |m|\leq 1 \,\, . \eqno(\eqn)$$ Since the absolute value of $m$ is
always less than or equal to $1$, toplighting always mutes a line
profile: hence our choice of ``muting'' for the name of the
toplighting effect on line profiles.
 
     We can consider a couple simple examples of muting by
toplighting.  First, consider $E=1/2$ and $R_{\rm cs}/R_{\rm ph}=2$.
Note that geometrically thin supernova atmospheres have not been
identified, and thus $R_{\rm cs}/R_{\rm ph}\ga2$ may be typical in
real supernovae.  With the given input values, $m=1/9$ and the
contrast factor of the line with respect to the continuum is reduced
by this factor in going from a standard case to a toplighting case.
The contrast factor of an absorption trough of a standard-case P-Cygni
line has absolute lower limit of $-1$ (i.e., zero flux).  Thus, even if
this lower-limit case were realized for a standard-case P-Cygni line,
the toplighting counterpart absorption would have absorption trough
depth of only $1/9$ of the continuum level.

     Next consider $E=\infty$ and $R_{\rm cs}/R_{\rm ph}=2$.  Here
$m=-1/7$.  Since $m$ is negative, the toplighting has produced a
flipped P-Cygni profile.  The standard-case P-Cygni absorption trough
would be turned into a toplighting-case emission peak with an upper
limit on the contrast factor of $1/7$.

     Figure~2 shows examples of P-Cygni files with $E$ values that
effectively span much of the $E$ parameter range.

     Since there is no upper limit
on the contrast factor of a standard-case P-Cygni emission peak,
{\it prima facie} it seems that a negative muting factor with large
absolute value could lead to negative observed flux in a corresponding
toplighting-case absorption component.
This does not happen of course.
As we showed above, the observed flux in the toplighting case is never
mathematically negative.
For another point of view, consider the following argument.
The standard-case P-Cygni line emission peak
is largest for an opaque line (i.e., one with $\tau=\infty$ everywhere
in the atmosphere) and large $R_{\rm cs}/R_{\rm ph}$.
For such a line, the emission peak
contrast factor can only increase with increasing
$R_{\rm cs}/R_{\rm ph}$ as $\sim \ln(R_{\rm cs}/R_{\rm ph})$ ($R_{\rm cs}$
considered here just as an outer boundary radius) (e.g., Jeffery
\&~Branch 1990, p.~189), but the absolute value of the muting factor
for large $R_{\rm cs}/R_{\rm ph}$
decreases for increasing $R_{\rm cs}/R_{\rm ph}$
like $(R_{\rm cs}/R_{\rm ph})^{-2}$.
Thus the muting factor always scales to prevent negative observed flux
from arising mathematically.

    The ratio of the monochromatic luminosities of the CSIR
and the photosphere (taken independently of each other) is
$$ \Gamma={ {L_{\rm cs}}\over{L_{\rm ph}}} ={ { 4\pi \left(2\pi R_{\rm
cs}^{2} I_{\rm cs}\right) } \over{ 4\pi \left(\pi R_{\rm ph}^{2}
I_{\rm ph}\right) } } = 2E\left({ {R_{\rm cs}}\over{R_{\rm ph}}
}\right)^{2} \,\, , \eqno(\eqn)$$ where the factor of $2$ accounts for
the fact that both hemispheres of the CSIR region contribute to flux
in any given direction.  The ratio $\Gamma$, in fact, appears in the
denominator of the muting factor formula, equation~(\showeqn -3).  If
we use equation~(\showeqn -0) to eliminate $E$ from equation~(\showeqn
-3), we obtain
$$ m={ { 2(R_{\rm cs}/R_{\rm ph})^{2} - \Gamma } \over { 2(R_{\rm
cs}/R_{\rm ph})^{2} - \Gamma +2(R_{\rm cs}/R_{\rm ph})^{2}\Gamma } }
\,\, . \eqno(\eqn)$$ Now $m$ as a function of $\Gamma$ monontonically
decreases with $\Gamma$ from $m=1$ at $\Gamma=0$ to a minimum
$m=-1/[-1+2(R_{\rm cs}/R_{\rm ph})^{2}]$ at $\Gamma=\infty$ (the only
stationary point).  It is clear that $\Gamma$ will have to be large in
some sense in order to obtain strong muting.  For the sake of
definiteness say $m\leq1/2$ is ``strong'' muting.  Then
$$ \Gamma\geq\Gamma\left(m={{1}\over{2}}\right)= { { 2(R_{\rm
cs}/R_{\rm ph})^{2} } \over { 2(R_{\rm cs}/R_{\rm ph})^{2} + 1 } }
\eqno(\eqn)$$ is required for strong muting.
Since $R_{\rm cs}/R_{\rm ph}\geq1$ is required physically, a necessary,
but not
sufficient, condition for strong muting is $\Gamma\geq2/3$.  If, as
suggested above, $R_{\rm cs}/R_{\rm ph}\ga2$ for supernovae, then a
necessary, but not sufficient, condition for strong muting in supernovae
is $\Gamma\ga8/9$.  Consequently, only those supernovae whose
monochromatic luminosities are strongly enhanced by circumstellar
interaction will have line profiles that are strongly muted by
toplighting.

\section{Picturesque Description}

     To complement the mathematical description of the
radiative transfer in the standard and toplighting cases of our 
simple model, we present here a picturesque description.

     First consider the standard case.  We use $R_{\rm cs}$ just as an
outer boundary of the atmosphere in this case.  If no line is present,
the photons emitted by the photosphere just escape to infinity and the
flux in any direction is just $I_{\rm ph}F_{0}$: this is the continuum
emission.  Adding a resonance-scattering line to the atmosphere has
the overall effect of reducing the wavelength-integrated emission of
the supernova in the wavelength interval that the line can affect.
This interval is $(\lambda_{0}+\Delta\lambda_{\rm
min},\lambda+\Delta\lambda_{\rm max})$, where
$$ \Delta\lambda_{\rm min}=-\lambda_{0}{ {R_{\rm cs}}\over{ct}}
\qquad{\rm and}\qquad
   \Delta\lambda_{\rm max}= \lambda_{0}{ {\sqrt{R_{\rm cs}^{2}-R_{\rm
ph}^{2}} }\over{ct}} \,\, .\eqno(\eqn)$$ The reason for the loss in
wavelength-integrated
emission is that the line, scattering isotropically, will scatter
some photons back to the photosphere where in our simple model they
are simply absorbed.  In a more realistic model, the photons absorbed
by the photosphere are a feedback that helps determine the
photospheric state.  The particular observer-directed photons which
are absorbed are those scattered toward the observer in the occulted
region: they simply hit the photosphere as they head toward the
observer.

     Because of the homologous expansion, photons continuously
redshift in the comoving frame of the atmosphere.  Thus, photons
emitted by the photosphere at or redward of $\lambda_{0}$ escape the
atmosphere without scattering.  (Note formally they can scatter at
$\lambda_{0}$ at the point of emission on the photosphere, but this
effect is assumed accounted for in specifying $I_{\rm ph}$, the
constant photospheric specific intensity.)  Thus the line scatters
photons that before scattering are blueward of $\lambda_{0}$ in
observer-frame wavelength.  The line scattering does not change a
photon's comoving frame wavelength (not at all in our simple model and
not significantly in reality), but by changing its propagation
direction relative to the matter flow it does change its
observer-frame wavelength.  Consider photons scattered toward a
distant observer.  If the scattering occurs from matter moving away
from, mainly perpendicular to, or toward the observer, then the
scattered photons in the observer frame have wavelengths redward of
$\lambda_{0}$, near $\lambda_{0}$, or blueward of $\lambda_{0}$,
respectively.

       From the proceeding remarks, it is clear that the observer
receives all the photons emitted by the photosphere into the line-of-sight
at and redward of $\lambda_{0}$ plus the unscattered photons
from blueward of $\lambda_{0}$.  In addition there are photons
scattered into the line-of-sight by the line from the whole wavelength
interval
$(\lambda_{0}+\Delta\lambda_{\rm min},\lambda+\Delta\lambda_{\rm max})$.
The scattering component is strongest near $\lambda_{0}$ where the
resonance planes for scattering are largest and they come closest to
(and even touch) the photosphere where the source function and usually
the scattering opacity are largest.  (Because of the nature of
homologous expansion all the photons scattered toward the observer
from a plane perpendicular to the line-of-sight have the same
observer-frame wavelength: see \S~2.)  These planes are near $z=0$.
The scattering component grows progressively weaker as one moves away
from near $\lambda_{0}$: i.e., as the resonance planes get farther
from $z\approx0$.  This scattering behavior results in the P-Cygni
profile emission feature with its peak near $\lambda_{0}$.

        The scattering of photons out of the line-of-sight
from blueward of $\lambda_{0}$ causes a flux deficiency relative to the
continuum:  this is the P-Cygni profile blueshifted absorption.
As we argued above the wavelength-integrated emission in the wavelength
interval
$(\lambda_{0}+\Delta\lambda_{\rm min},\lambda+\Delta\lambda_{\rm max})$
is less than in the line's absence.
Consequently, the P-Cygni absorption feature will be larger than the
P-Cygni emission feature.
(This ratio of feature size is not necessarily obtained if the line
is not a pure resonance scattering line.)

        Now consider the toplighting case.
First, imagine that there is no photosphere or line;
there is just the radiating, optically transparent,  spherical shell CSIR.
The flux in any direction is $2I_{\rm cs}G_{0}$ at all wavelengths.
Now add a line to the atmosphere enclosed by the CSIR.
The line has, in fact, no effect on the flux emission. 
The comoving-frame radiation field at the line center wavelength at any
point inside the atmosphere is isotropic.
The line scattering is isotropic.
In the comoving frame, an isotropic field isotropically scattered is unchanged.
Since the comoving frame radiation field is unchanged,
the observer-frame radiation field is also unchanged.

     Now remove the line, but add a non-emitting (but, of course,
opaque) photosphere.  The flux in all directions at all wavelengths is
now $I_{\rm cs}\left(2G_{0}-F_{0}\right)$.  The photosphere just acts
as a net absorber.  But if one now adds a line, the line will scatter
some photons directed toward the photosphere into directions leading
to escape.
The result is that the wavelength-integrated flux
in the wavelength interval $(\lambda_{0}+\Delta\lambda_{\rm
min},\lambda+\Delta\lambda_{\rm max})$ is increased by the addition of
the line.

      Again one has to consider how the line scattering shifts the
observer-frame wavelength.  The non-emitting photosphere causes there
to be a ``cone of emptiness'' in the radiation field converging at any
point in the atmosphere.  This reduces the line source function at
every point in the atmosphere from the no-photosphere case.  Thus,
observer-directed beams in the limb region of the
atmosphere that interact with the line must be reduced from the
no-photosphere case.  The closer the resonance point for a beam is to
the $z=0$ location, the greater the effect of the ``cone of
emptiness'', and the weaker the emergent beam will be.  Beams that do
not interact with the line will be unchanged from the no-photosphere
case.  The upshot is that centered on $\lambda_{0}$ there will be flux
deficit relative to the continuum: an absorption feature around the
line center wavelength.

      On the other hand, beams from the photodisk region (i.e., the
non-limb part of the atmosphere on the near side of the photosphere)
are enhanced by the line scattering.  Without the line, the photodisk
specific intensity is just $I_{\rm cs}$ from the near hemisphere of
the CSIR; the beams from the far hemisphere of the CSIR are occulted
by the photosphere.  But with the line there is scattering into the
line-of-sight in the photodisk region.  Consequently, there is an
enhancement in flux over the continuum: i.e., a flux emission feature.
Since the photodisk region is moving toward the observer this flux
emission feature is blueshifted from $\lambda_{0}$.

       One thus finds that toplighting with a non-emitting photosphere
gives rise to a flipped P-Cygni line with a blueshifted emission and
an absorption
centered near the line center wavelength.  From the argument about
wavelength-integrated flux above, we see that the blueshifted emission
must be larger than the absorption trough centered near line center wavelength.

       In a general toplighting case the photosphere emits, and so
there is a competition between P-Cygni and flipped P-Cygni line
formation.  Clearly in most supernovae the competition is won by
P-Cygni line formation.  For flipped P-Cygni line formation to win,
$E=I_{\rm cs}/I_{\rm ph}$ must be greater than 1 (see \S~4).

\section{Discussion}

The treatment of toplighting in supernova resonance-scattering line
formation has been presented here in its simplest form for its
heuristic value.  For example, we have not discussed how the
toplighting might change the radial dependence of the line optical
depth $\tau(r)$, and we have not considered the possibility that the
CSIR reflects supernova light back into the line--forming region.  The
toplighting version of the SYNOW code that we (Blaylock et~al.  1999,
in preparation) are using to analyze the spectra of SN~1998S allows
for an angular dependence of the radiation from an optically thin
circumstellar shell and for a wavelength dependence of the
photospheric and CSIR specific intensities.  Including toplighting
allows us to obtain improved fits to the observed spectra of SN~1998S
and shows that some of the 
P-Cygni line profiles in the early-time SN~1998S may be flipped.
(It is not easy to be sure of the flipped profiles
because of the numerous superimposed circumstellar features.)
Lentz et~al. (1999, in preparation) find that allowing for toplighting
in detailed NLTE calculations also leads to improved fits.

Among core-collapse events, SN~1998S has been uniquely well observed
at ultraviolet wavelengths, but physically it is not an exceptional
case.  Line profiles in other circumstellar-interacting core-collapse
events presumably also are affected by toplighting.
But as we showed
in \S~4, only those supernovae whose monochromatic luminosities are
strongly enhanced by circumstellar interaction will have line profiles
that are strongly muted by toplighting.

Examples of supernovae with relatively strong circumstellar interaction
(as known from relatively strong radio emission) are
SN~1979C, SN~1980K and SN~1993J.
All these supernovae showed rather featureless UV spectra in the
$\sim 1800$--$2900\,$\AA\ region in comparison to supernovae known not
to have had strong circumstellar interaction (Jeffery~et~al. 1994).
Perhaps a UV-peaked toplighting continuum was muting the UV spectra
of SN~1979C, SN~1980K and SN~1993J.

Spectropolarimetry and nebular-phase line profiles indicate that
SN~1998S and its circumstellar shell were not spherically symmetric
(Leonard et~al. 1999) as has been assumed here, and core-collapse
events in general appear to be asymmetric (Wang~et al.~1996).
Eventually, asymmetric toplighting will have to be taken into account
in spectrum calculations.

This work was supported by NASA grant NAG5--3505, NASA grant
GO-2563.001 to the SINS group from the Space Telescpe Science
Institute, which is operated by AURA under NASA contract NAS~5--26555,
and Middle Tennessee State University.

\clearpage

\begin {references}

\reference{} Blaylock, M., et al. 1999, in preparation

\reference{} Garnavich, P. M.,  et al. 1999, in preparation

\reference{} Hatano, K., Branch, D., Fisher, A., Baron, E., 
             Filippenko, A. V. 1999, ApJ, in press, astro-ph/9903333

\reference{} Jeffery, D. J., \& Branch, D. 1990, in Jerusalem Winter
             School for Theoretical Physics, Vol.~6, Supernovae,
             ed.~J.~C.~Wheeler, T.~Piran, \&~S.~Weinberg (Singapore:  World
             Scientific), 149 

\reference{} Jeffery, D. J.,~et al.
             1994, ApJ, 421, L27

\reference{} Jeffery, D. J., \& Vaughan, T. E. 1999, in preparation

\reference{} Lentz, E. J.,  et al. 1999, in preparation

\reference{} Leonard, D. C., Filippenko, A. V., Barth, A. J., Matheson, T.
             1999, ApJ, submitted, astro-ph/9908040

\reference{} Mihalas, D. 1978, Stellar Atmospheres (San Francisco:  Freeman)

\reference{} Millard, J., et al. 1999, ApJ, in press, astro-ph/9906496

\reference{} Rybicki, G. B., \&~Hummer, D. G. 1978, ApJ, 219, 654

\reference{} Wang, L., Wheeler, J. C., Li., Z. W. \&~Clocchiatti, A. 1996,
            ApJ, 467, 435

\end{references}

\clearpage

\begin{figure} 
\figcaption{A simple model of supernova-line formation in the
presence of toplighting.}
\end{figure}

\begin{figure} 
\figcaption{Supernova resonance-scattering line profiles in the
presence of toplighting for $R_{\rm cs}/R_{\rm ph}=2$}
\end{figure}

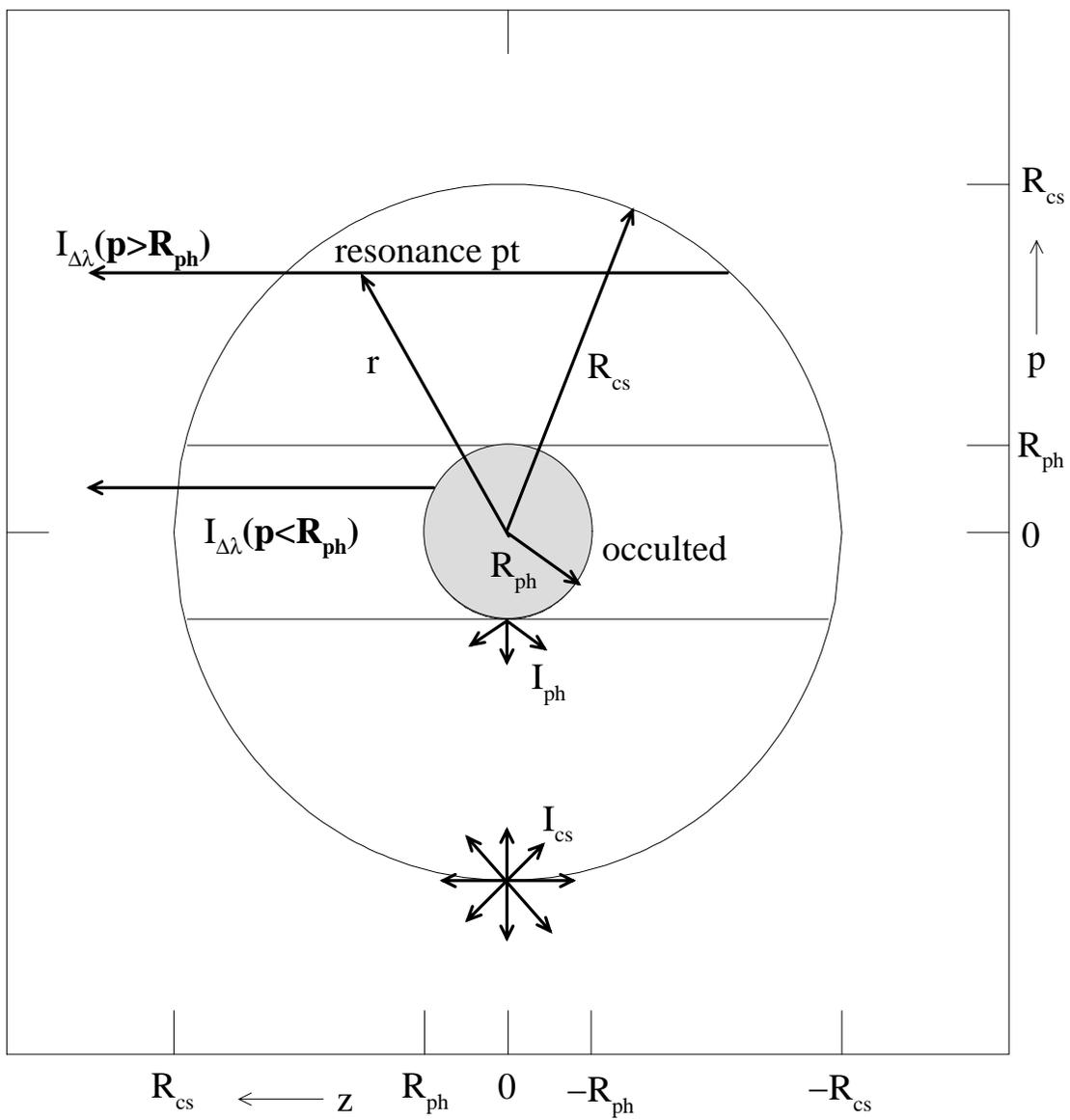
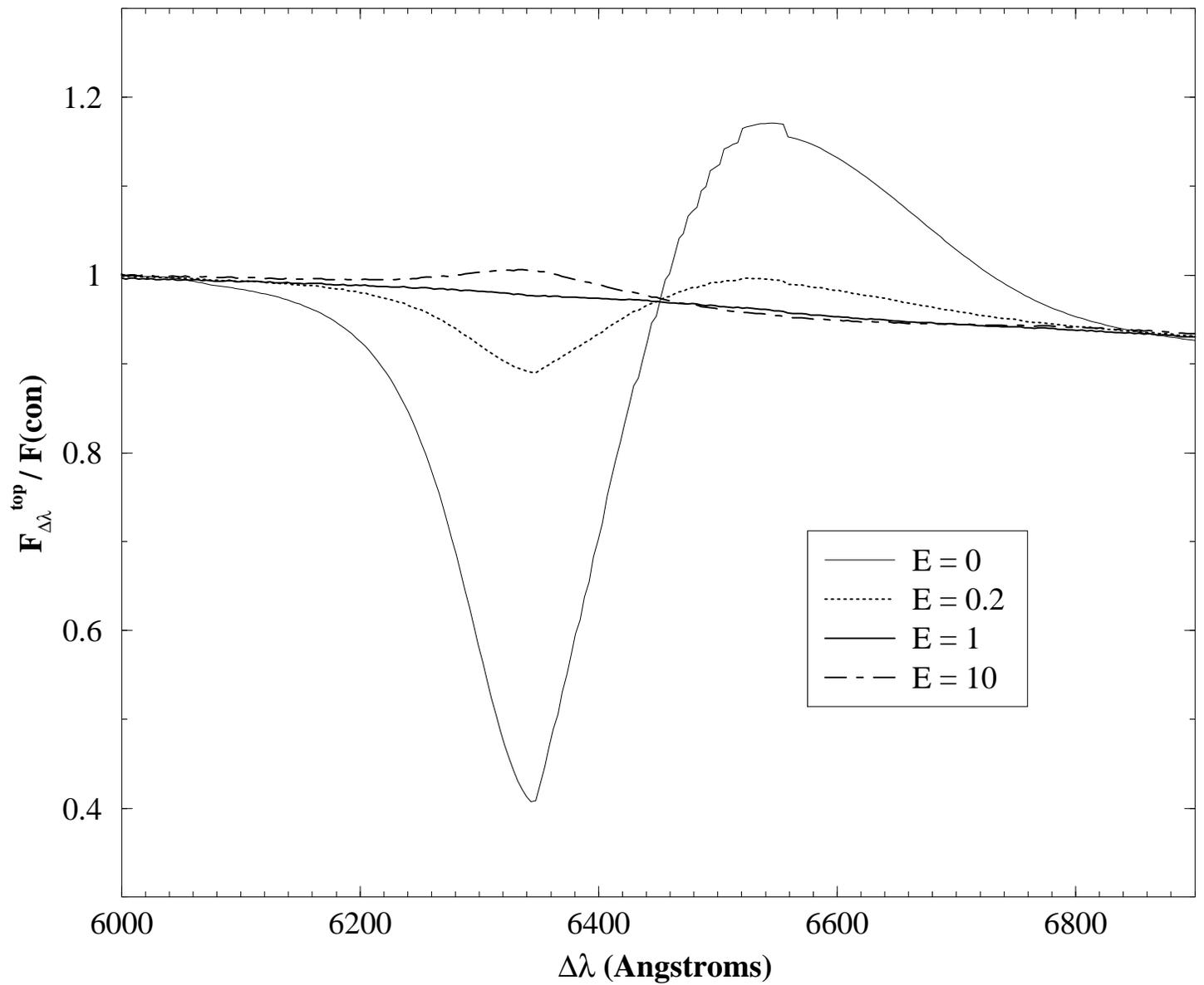
\end{document}